\documentclass[12pt,preprint]{aastex}

\makeatletter
\renewcommand{\fps@figure}{thbp}
\renewcommand{\fps@table}{thbp}

\makeatother

\newcommand{\kms}{{\,\rm km\,s}^{-1}} 
 
\newcommand\dm{\mbox{$.\!\!^{\mathrm d}$}}%
\def\la{\mathrel{\hbox{\rlap{\hbox{\lower4pt\hbox{$\sim$}}}\hbox{$<$}}}}
\def\sun{\hbox{$\odot$}}
\def\as{\hbox{$.\!''$}}

\begin{document}
\title{ The Cepheid Distance to NGC\,5236 (M\,83) with the 
VLT\footnote{Based on observations collected at the UT1 of the 
Very Large Telescope, which is operated by the European
Southern Observatory.} \\}

\author{Frank Thim\footnote{Visiting fellow to NOAO}, G.A. Tammann} 
\affil{Astronomisches Institut der Universit\"at Basel \\ 
Venusstrasse 7, CH-4102 Binningen, Switzerland} 
\author{A. Saha, A. Dolphin} 
\affil{National Optical Astronomy Observatories \\
950 North Cherry Ave., Tucson, AZ 85726} 
\author{Allan Sandage}
\affil{The Observatories of the Carnegie Institution of Washington \\
813 Santa Barbara Street, Pasadena, CA 91101} 
\author{E. Tolstoy} 
\affil{Kapteyn Instituut, Rijksuniversiteit Groningen \\ 
   NL-9700AV Groningen, Netherlands}
\author{Lukas Labhardt} 
\affil{Astronomisches Institut der Universit\"at Basel \\ 
Venusstrasse 7, CH-4102 Binningen, Switzerland} 

\begin{abstract}

Cepheids have been observed in NGC\,5236 (M\,83) using 
the ANTU (UT1) 8.2 meter telescope of the ESO VLT with FORS1.
Repeated imaging observations have been made between January 2000 and 
July 2001. Images were obtained on 34 epochs
in the $V$ band and on 6 epochs in the $I$ band. 
The photometry was made with the ROMAFOT reduction package
and checked independently with DoPHOT and a modified
version of HSTphot. Twelve Cepheid
candidates have periods ranging between 
12 and 55 days. The dereddened 
distance modulus is adopted to be $(m-M)^{0} = 28.25 \pm 0.15$,
which corresponds to a distance of $4.5 \pm 0.3$ Mpc.
The Cepheid distance of NGC\,5253 has been
rediscussed and strengthened by its SN~1972E. The mean 
distance of $(m-M)^{0} = 28.01 \pm 0.15$ 
(based on SN~1972E) shows the galaxy
to be a close neighbor of M\,83, suggesting that the two
galaxies may have interacted in the past and thus
possibly explaining the amorphous morphology of
NGC\,5253. The distance difference between M\,83 and 
NGC\,5253 is only ($0.5 \pm 0.4$) Mpc. The 
{\it projected\/} distance is only 
$\sim$ 0.15 Mpc. 

M\,83 is the principal member of the nearby M\,83 group 
containing also, besides NGC\,5253, several dwarf members,
for five of which TRGB distances are available 
\citep[][\aap, 385, 21]{Karachentsev:etal:2002a}.
The adopted group distance of $(m-M)^{0} = 28.28 \pm 0.10$
($4.5 \pm 0.2$ Mpc) together with its mean recession velocity 
of $v_{\rm LG} = 249 \pm 42$ km s$^{-1}$ shows again 
the extreme quietness of the local (1 Mpc to 10 Mpc) expansion 
field. M\,83 fits onto the local mean Hubble flow line of the 
velocity-distance relation (with H$_0 \sim 60$)
with no significant deviation, 
supporting the earlier conclusion 
that the local velocity expansion field is remarkably cold 
on a scale of 10 Mpc, contrary to the predictions of the 
simplest cold dark matter model for large scale structure. The 
role of a cosmological constant has been invoked 
as a possible solution in providing a nearly uniform force 
field everywhere in the presence of a lumpy galaxy 
distribution.

\end{abstract}


\section{Introduction} 

Galaxy distances are the basis of much of extragalactic astronomy and a 
central theme in cosmology. There is a wide, yet poorly tested
consensus that the most 
reliable extragalactic distances come from the period-luminosity (P-L) 
relation of Cepheid variable stars. If observed in two passbands
(V,I), their absorption-corrected distances can be determined. 
The zero point of the P-L 
relation is usually based on an adopted distance of 
LMC, whose distance is secure to $\pm 0.1$
from a number of distance indicators
\citep[cf.\ compilations by][]{Federspiel:etal:98, Gibson:2000,
Tammann:etal:2003a}.

The spiral galaxy M\,83 (NGC\,5236) 
($\alpha_{2000}$=$13^{\rm h}37^{\rm m}01^{\rm s}$, 
$\delta = -29^{\circ}51^{\rm m}59^{\rm s}$),
classified as SBc(s)II in the RSA 
\citep{Sandage:Tammann:87}, is the principal member
of a small galaxy group comprising in addition nine
probable dwarf members \citep{Karachentsev:etal:2002a}.
NGC\,5253 is a almost certain member of the 
group. The group is, however, distinct from the NGC\,5128
(Cen A) group which has in addition to NGC~4945 probable
dwarf members; this group lies about 12$^0$ away from M\,83,
has a smaller mean distance, and a lower mean redshift 
according to \citet{Karachentsev:etal:2002a}.

The distance of M\,83 is particularly interesting for two 
reasons. 1) NGC\,5253 lies only $\sim$ 2$^0$ from M\,83.
It is a prototype of the amorphous class 
\citep{Sandage:Brucato:1979}. It has been suggested that 
amorphous galaxies are the result of gravitational 
interaction \citep{Krienke:Hodge:1974, Hogg:etal:1998},
but in the case of NGC\,5253 no interacting partner has 
yet been proposed. Since the distance of NGC\,5253
is well known from its Cepheids and SNIa 1972E, a good 
distance of M\,83 becomes highly desirable for
comparison. 2) The distances to local galaxies
known at present suggest that the local 
flow pattern of the Hubble 
expansion field is unexpectedly quiet \citep{Sandage:etal:1972,
Sandage:1986, Ekholm:etal:2001, Tammann:etal:2001,
Karachentsev:etal:2002b, Karachentsev:etal:2003}.
M\,83 provides a valuable additional local datum with which
to map the local velocity field. 

From the beginning of the modern mapping of the local 
expansion field, a principal objective has been to determine the 
velocity dispersion about the mean Hubble flow. 
\citet{Hubble:Humason:1931, Hubble:Humason:1934}
had early estimated that the velocity 
dispersion about the linear velocity-distance relation was $\la$ 200 
km s$^{-1}$. As the estimates of relative distances became better, the 
value steadily decreased. By 1972 \citet*{Sandage:etal:1972}
could measure an upper limit of $\sigma(\Delta v) \sim $
100 km s$^{-1}$. This was reduced to $\sim$ 50 km s$^{-1}$ by 
\citet{Sandage:Tammann:1975} in Paper V of their Hubble Constant 
Steps series. This low value has been confirmed often thereafter 
by others \citep[e.g.][]{Sandage:1975, Sandage:1986,
Tammann:Kraan:1978, Ekholm:etal:2001}. Also, the 
many new Cepheid distances to very local galaxies just outside 
the Local Group in the programs by Hoessel and Saha and their 
collaborators and other groups 
[see \citet{Mateo:1998} for extensive 
references]
confirms that $\sigma(\Delta v) < $ 60 km s$^{-1}$ 
for distances up to 7 Mpc beyond the Local Group.  

     Because this extremely low value contradicts the prediction 
of the simplest cold dark matter model for the formation of large 
scale structure by at least a factor of 5, \citep[cf.][]{Davis:Peebles:1983,
Davis:etal:1985, Ostriker:1993, Governato:etal:1997, 
Bertschinger:1998}, continued measurements of the quietness of 
the Hubble flow over distance scales within 10 Mpc are crucial. 
The galaxy group with M\,83 as a member at a mean distance of $\sim$ 4.5 
Mpc (determined here) is of special importance. 

\section{Observations} 

\subsection{The instruments} 

Repeated imaging of M 83 has been made with the ESO Very 
Large Telescope (VLT)
Unit Telescope 1 (ANTU) at Paranal Observatory in Chile. 
The instrument used was FORS1 (FOcal Reducer and low dispersion 
Spectrograph) with a 2048$\times$2048 Tektronix CCD with 
$24 \mu$m pixels.
Two different spatial resolutions can be selected.
We used the standard resolution collimator, which delivers a lower 
resolution but larger field-size. 
This provides a field of view of $6\farcm8 \times 6\farcm8$ and 
a pixel scale of $0\farcs2/$pixel. The large collecting area makes 
FORS1 superior to HST/WFPC2 for the detection of Cepheids 
in NGC\,5236. For the determination of the internal absorption, 
observations with Bessel $V$- and $I$-band filters 
\citep{Szeifert:Boehnhardt:2001}
have been made.

\subsection{The Data} 

A field northwest of the center of M\,83 was chosen. Images of the
center of M\,83 would be too crowded for high quality photometry.
A 20 x 20 arcmin$^2$ field of the Digitized Sky
Survey at the position of M 83 is shown in Fig.~\ref{fig1}. The square
box shows the FORS1 $6\farcm8 \times 6\farcm8$ field of view. 
There are in total 34 epochs in the $V$ passband and 6 epochs
in the $I$ passband over a period of 1.5 years, from 2000 January 04
to 2001 July 23.  Each individual epoch consists of 2-4 
subexposures having exposure times between 400 and 600 seconds.
A journal of observations is given in Table~\ref{tab1}. The seeing
is almost always sub-arcsecond and for some epochs exceptionally good 
($\le$ 0.5 arcsec). Such exceptional seeing conditions are 
required at least in one $V$ and one $I$ image to get 
accurate stellar positions
which are used to disentangle the stellar photometry on images 
of lower seeing quality. The VLT image of the
epoch with the best seeing, i.e. V\_17, is shown in Fig.~\ref{fig2}. 



{\clearpage
\begin{deluxetable}{lcccccc}
\tabletypesize{\footnotesize}
\tablecaption{Observation Log. }
\tablewidth{0pt}
\tablehead{
\colhead{Archival Image ID\tablenotemark{a}} &
\colhead{HJD\tablenotemark{b}} & 
\colhead{Exp. Time (s)} &
\colhead{Filter} &
\colhead{Seeing\tablenotemark{c}} &
\colhead{Airmass\tablenotemark{d}} & 
\colhead{Image ID\tablenotemark{e}}}
\startdata
 FORS.2000-01-05T08:02 .. 08:20 & 51548.345 & 3 x 500 & V & 0.55 .. 0.68 &  1.34 & V\_01 \\ 
 FORS.2000-01-09T08:22 .. 08:40 & 51552.360 & 3 x 500 & V & 0.61 .. 0.67 &  1.21 & V\_02 \\
 FORS.2000-01-18T07:48 .. 08:06 & 51561.336 & 3 x 500 & V & 0.57 .. 0.67 &  1.20 & V\_03 \\ 
 FORS.2000-02-04T08:00 .. 08:18 & 51578.343 & 3 x 500 & V & 0.60 .. 0.66 &  1.04 & V\_04 \\
 FORS.2000-02-04T08:28 .. 09:00 & 51578.367 & 4 x 600 & I & 0.56 .. 0.59 &  1.01 & I\_01 \\
 FORS.2000-02-04T09:12 .. 09:30 & 51578.394 & 3 x 500 & V & 0.51 .. 0.55 &  1.00 & V\_05 \\ 
 FORS.2000-02-05T09:03 .. 09:21 & 51579.386 & 3 x 500 & V & 1.21 .. 1.39 &  1.00 & V\_06 \\ 
 FORS.2000-02-08T05:58 .. 06:26 & 51582.261 & 4 x 500 & V & 0.58 .. 0.79 &  1.29 & V\_07 \\
 FORS.2000-03-12T06:06 .. 06:25 & 51615.262 & 3 x 500 & V & 0.75 .. 0.88 &  1.01 & V\_08 \\
 FORS.2000-03-12T06:35 .. 06:53 & 51615.281 & 3 x 500 & V & 0.62 .. 0.75 &  1.00 & V\_09 \\ 
 FORS.2000-03-12T07:04 .. 07:34 & 51615.301 & 3 x 500 & V & 0.57 .. 0.64 &  1.00 & V\_10 \\
 FORS.2000-03-12T07:33 .. 07:51 & 51615.320 & 3 x 500 & V & 0.64 .. 0.73 &  1.02 & V\_11 \\
 FORS.2000-03-12T08:01 .. 08:19 & 51615.340 & 3 x 500 & V & 0.70 .. 0.74 &  1.04 & V\_12 \\
 FORS.2000-03-12T08:29 .. 08:47 & 51615.359 & 3 x 500 & V & 0.81 .. 0.82 &  1.08 & V\_13 \\ 
 FORS.2000-03-12T08:57 .. 09:29 & 51615.383 & 4 x 600 & I & 0.83 .. 0.94 &  1.16 & I\_02 \\
 FORS.2000-03-13T03:37 .. 04:06 & 51616.160 & 4 x 500 & V & 0.62 .. 0.66 &  1.31 & V\_14 \\
 FORS.2000-03-13T04:16 .. 04:27 & 51616.184 & 2 x 600 & I & 0.69 .. 0.83 &  1.21 & I\_03 \\
 FORS.2000-03-15T06:33 .. 06:51 & 51618.281 & 3 x 500 & V & 1.14 .. 1.75 &  1.00 & V\_15 \\
 FORS.2000-03-31T06:00 .. 06:19 & 51634.256 & 3 x 500 & V & 0.52 .. 0.61 &  1.01 & V\_16 \\
 FORS.2000-04-05T06:16 .. 06:34 & 51639.264 & 3 x 500 & V & 0.35 .. 0.40 &  1.03 & V\_17 \\
 FORS.2000-05-06T04:54 .. 05:26 & 51670.210 & 4 x 500 & V & 0.69 .. 0.93 &  1.09 & V\_18\tablenotemark{f} \\
 FORS.2000-07-01T01:21 .. 01:36 & 51726.057 & 3 x 400 & V & 0.88 .. 1.04 &  1.09 & V\_19 \\
 FORS.2000-07-07T23:06 .. 23:21 & 51732.964 & 3 x 400 & V & 0.51 .. 0.57 &  1.00 & V\_20 \\
 FORS.2000-07-07T23:29 .. 23:57 & 51732.983 & 3 x 600 & I & 0.46 .. 0.49 &  1.01 & I\_04 \\
 FORS.2000-07-26T00:11 .. 00:26 & 51751.008 & 3 x 400 & V & 0.92 .. 1.00 &  1.16 & V\_21 \\
 FORS.2000-07-30T23:27 .. 23:42 & 51755.981 & 3 x 400 & V & 0.95 .. 1.05 &  1.10 & V\_22 \\
 FORS.2000-08-07T23:37 .. 23:52 & 51763.985 & 3 x 400 & V & 1.00 .. 1.14 &  1.21 & V\_23 \\
 FORS.2001-02-17T07:07 .. 07:23 & 51957.303 & 3 x 400 & V & 1.06 .. 1.65 &  1.04 & V\_24 \\
 FORS.2001-03-22T04:26 .. 04:41 & 51990.191 & 3 x 400 & V & 0.54 .. 0.89 &  1.09 & V\_25 \\
 FORS.2001-03-27T06:39 .. 06:54 & 51995.280 & 3 x 400 & V & 0.70 .. 0.75 &  1.02 & V\_26 \\ 
 FORS.2001-05-17T00:37 .. 00:52 & 52046.026 & 3 x 400 & V & 0.61 .. 0.83 &  1.11 & V\_27 \\
 FORS.2001-05-20T23:17 .. 23:42 & 52049.975 & 4 x 400 & V & 0.64 .. 0.77 &  1.26 & V\_28\tablenotemark{f} \\
 FORS.2001-05-26T03:17 .. 03:32 & 52055.135 & 3 x 400 & V & 1.39 .. 1.58 &  1.05 & V\_29 \\
 FORS.2001-05-28T02:16 .. 02:31 & 52057.096 & 3 x 400 & V & 0.66 .. 0.79 &  1.01 & V\_30 \\
 FORS.2001-06-16T01:54 .. 02:16 & 52076.081 & 3 x 600 & I & 0.47 .. 0.53 &  1.06 & I\_05 \\ 
 FORS.2001-06-20T23:47 .. 00:08 & 52080.995 & 3 x 600 & I & 0.45 .. 0.56 &  1.00 & I\_06 \\ 
 FORS.2001-06-26T23:23 .. 23:38 & 52086.975 & 3 x 400 & V & 0.49 .. 0.53 &  1.00 & V\_31  \\ 
\tableline
\\
 FORS.2001-07-17T00:05 .. 00:20 & 52107.003 & 3 x 400 & V & 0.48 .. 0.55 &  1.07 & V\_32 \\ 
 FORS.2001-07-20T00:25 .. 00:40 & 52110.019 & 3 x 400 & V & 0.68 .. 0.83 &  1.14 & V\_33 \\
 FORS.2001-07-23T23:30 .. 23:45 & 52113.980 & 3 x 400 & V & 0.73 .. 1.06 &  1.06 & V\_34 \\
\enddata
\tablenotetext{a}{The image ID's as used in the Archive, i.e. the instrument 
prefix followed by a timestamp. The timestamp in this table is year,
month, date, hour and minute (UTC). Instead of listing all images 
the range of the subexposures are shown.}
\tablenotetext{b}{Heliocentric Julian date - 2400000.5 at midexposure.}
\tablenotetext{c}{The range of FWHM seeing conditions for all subexposures.}
\tablenotetext{d}{The average airmass.}
\tablenotetext{e}{The name of the combined images. Each combined image
consists of 2-4 subexposures.}
\tablenotetext{f}{Not all subimages have been used to create the combined 
image, i.e. FORS.2000-05-06T05:12 and FORS.2001-05-20T23:17 
have not been used. FORS.2000-05-06T05:12 has been exposed twice,
FORS.2001-05-20T23:17 has been taken above the requested image quality 
and was repeated.}
\label{tab1}
\end{deluxetable}
}
\clearpage

\section{Data Processing and Photometry} 

In order to gain control over the errors introduced by individual
photometry procedures and calibrations, the data have been 
processed independently by us using three different photometry 
software programs with three independent calibrations:
the ROMAFOT package \citep{buo83}
as implemented in ESO-MIDAS (European Southern Observatory -
Munich Image Data Analysis System) \citep{ESO:1992}
which was developed for crowded fields, 
DoPHOT \citep{schec93} as modified by one of us (A. Saha)
for ground based images, 
and a modified version of HSTphot
\citep{Dolphin:etal:2002}. 
ROMAFOT was
used for the photometry of all epochs including the variable search,
whereas DoPHOT and HSTphot have been used only to process the best
$V$ and $I$ images to compare the calibration with respect to each
other. A detailed discussion of photometry with ROMAFOT including
artificial star experiments and undersampling can be 
found in \citet{Thim:2001}, previous applications of ROMAFOT  
in \citet{Saha:etal:2001a}, \citet{Tammann:etal:2001} and 
\citet{Thim:2000}.

\subsection{Data Reductions}
The FORS1 direct-imaging observations have been performed in 
service mode by Paranal Science Operations staff.
We used the pipeline reduced images which were provided by
Garching Quality Control group which were bias-subtracted and
flat fielded for us.
The raw images have been reduced 
independently and checked with the pipeline-produced images.
No significant differences were found.
A point to note is the gain setting. 
FORS1 has different gain settings. First, the gain 
is different for the four different quadrants (4-ports readout). 
Second, the gain setting is different for the science images and
for the standard star observations. The science images are 
always taken with low gain, the standard star observations 
are taken either with low or high gain. While the relative gain 
between the 4-quadrants
is pipeline-calibrated, the overall gain settings between
science and standard star images have to be corrected 
individually.

\subsection{Relative Photometry with ROMAFOT}

The two best images; V\_17 and I\_04, taken with a seeing
of $\leq 0\as50$,
were used as reference images. Attempts
to create an even deeper reference image by co-adding the 
images of various epochs were unsuccessful due to variable seeing
and non-negligible field shifts and field rotation. 

25 isolated, unsaturated stars 
were selected on V\_17 and I\_04 to establish the mean PSFs in $V$ 
and $I$. The PSF of each image was found not to vary 
significantly with position.

Stars on V\_17 and I\_04, brighter than a certain limit,
were then fitted
with the appropriate PSF and subtracted from the field. 
The procedure was repeated on the residual images in an iterative
process cutting at fainter and fainter limits.

The resulting list of stars - separately in $V$ and $I$ -
and their positions were used as a master list for all other
images. The master positions were transformed into 
positions on the individual images by means of a matching
algorithm. 

All stars of the master lists were searched on the images in
$V$ and $I$, respectively, and fitted with the mean PSF
of that image going stepwise, as above, to the
faintest possible stars. The mean PSF was determined 
using the same 25 stars, if possible, as above.

\subsection{Standardizing the Photometry}

The PSF fitted magnitudes were corrected
for atmospheric extinction. Since the Landolt standard
stars (see below) were observed at almost constant airmass, 
it was not possible to determine nightly extinction
coefficients. Mean extinction coefficients from
the VLT homepage have therefore been adopted.

The corrected PSF magnitudes of all $V$- and $I$-images
are still on
arbitrary zero points due to different exposure times
and seeing. The mean zero point shifts of each image with 
respect to the template images V\_17 and I\_04,
respectively, were determined, using all well fitted stars,
and then subtracted. Thus all PSF magnitudes are
converted to the same zero point. 

The next step is to convert the PSF magnitudes, 
which rely on the core of the PSF, to aperture
magnitudes by means of the additive aperture
corrections (AC). The AC was determined for
the images V\_17 and I\_04 by performing aperture photometry 
on all sufficiently isolated, unsaturated stars with flat 
growth curves. Only four stars in V\_17 and five
stars in I\_04 were accepted for the best mean ACs.
The random error of the adopted mean ACs is
$0\fm{02}$ in $V$ and $0\fm{01}$ in $I$. 
The value of the AC depends on the selected stars, the
systematic error of the AC is estimated using different 
samples of stars, i.e. the standard deviation of 
different solutions for the AC. Their systematic error 
is estimated to be $0\fm{05}$.

The aperture magnitudes are converted into 
instrumental magnitudes by adding the appropriate
AC to all stars of the various images.
The instrumental magnitudes 
were then transformed into standard magnitudes. The
transformation equations were determined from
13 Landolt standard stars, covering the color range
$-0.53 < (V-I) < 1.95$, in the fields Ru 152 and
PG 0231, which were observed together with
V\_17 and I\_04. The resulting equations are:
\begin{equation}
V = V_{\rm instr} + 0.05(\pm0.012) * (V_{\rm instr} - I_{\rm instr}) + 27.77(\pm0.007)
\end{equation}
\begin{equation}
I = I_{\rm instr} - 0.04(\pm0.013) * (V_{\rm instr} - I_{\rm instr}) + 26.90(\pm0.011), 
\end{equation}
$V_{\rm instr}$ and $I_{\rm instr}$ are the instrumental magnitudes, corrected
for extinction. The standard system is reproduced to within
$\pm0.01$ with a rms. error of $0\fm{02}$ in $V$ and in $I$.

\subsection{Comparing Results from the different Photometric Calibrations with 
ROMAFOT, DoPHOT, and HSTphot}

The combined V- and I-images with the best seeing, i.e. V\_17
and I\_04, were independently reduced with three different 
software programs (by three different members of our group). 

The ROMAFOT photometry is characterized in detail above. 
The DoPHOT photometry was carried out as described in 
\citet{Saha:etal:2002}. The photometry with the modified 
version of HSTphot was carried out as described in 
\citet{Dolphin:etal:2002}.
A comparison of the resulting
magnitudes in $V$ and $I$ is presented in 
Fig.~\ref{fig5} - ~\ref{fig7}. Note that
the apparent up-turn of the point distributions at the
faintest levels is simply due to the respective cutoffs of the
different magnitude systems, i.e. it is a bias due to
incompleteness. 

The aperture correction (AC) was determined independently
for each photometry by different members of our group with
a number of undisturbed stellar images; different stars were
used for the three different software packages. The resulting 
independent
photometric zero points agree to within 0.03 in $V$ (at $V$ = 20.0)
and $I$ (at $I$ = 19.0). We have decided to use the 
zero points in $V$ and $I$ of the ROMAFOT system throughout this paper 
because they lie in between the zero points of DoPHOT and 
HSTphot. The adopted zero points are estimated to be accurate 
to within $\pm 0\fm{03}$.

Fainter than the level of the adopted common zero point, i.e. at
$V >$ 20.0, $I > 19.0$, the photometric scales of the three
independent reduction procedures differ slightly. The three 
sets of $V$ magnitudes of about 250 stars with $V <$ 23.5
were compared. They reveal a scale difference between 
ROMAFOT and DoPHOT of $0\fm{013} \pm 0\fm{002}$ per 
magnitude, and between ROMAFOT and HSTphot of 
$0\fm{023} \pm 0\fm{001}$ per magnitude. The corresponding
scale difference in $I$, again from about 250 stars with 
$I <$ 22.5 is between ROMAFOT and DoPHOT 
$0\fm{015} \pm 0\fm{002}$ and between ROMAFOT and HSTphot
$0\fm{035} \pm 0\fm{001}$ per magnitude.
The sense of the
differences is that ROMAFOT is always fainter. We have
adopted an intermediate scale. This results in a mean
correction of all ROMAFOT magnitudes by $0\fm{009}$ 
per magnitude in $V$ and $0\fm{017}$ per magnitude in $I$.
These corrections have been individually applied to all
Cepheid magnitudes in Table~\ref{tab2} below. The 
maximum corrections of the faintest Cepheid magnitudes
amounts to $-0\fm{04}$ in $V$ and $-0\fm{07}$ in $I$.




Photometry at these levels of faintness and crowding are 
extremely vulnerable to how the background sky is measured. 
The three programs have some differences in their respective 
prescriptions for estimating background. 
ROMAFOT and DoPHOT fit the background as part of the PSF fitting. 
HSTphot estimates background from the statistics of pixels surrounding 
the object. There are pros and cons to both approaches, and a philosophical 
discussion is beyond the scope of this paper. The differences in background 
that result from such procedural differences are small, but even systematic 
differences of a few analog to digital units (ADUs)
(of the order of sampling errors) can produce 
noticeable differences in the measured magnitudes of faint stars. If 
program A measures sky systematically lower than program B, the stars 
measured by A will be brighter than by B. These effects are negligible for 
brighter stars, but as one approaches the detection limit, the differences 
increase. 

There is another, perhaps more significant way in which background 
measurements can be different.
All three of the programs have a procedure where objects 
already identified are subtracted before a background
estimate is made. However, the procedure by which objects are identified
are different, and thus the exact list of objects that are subtracted from 
a given patch of sky will be different. Again, this results in 
systematic differences in measured background, and manifests as differences 
in scale for measured magnitudes near the faint end. 
Due to the nature of luminosity 
functions, both of stars, and of background galaxies, the problem gets 
more acute as one goes fainter. 
     
In the face of this, there is no clear cut `correct' prescription. Instead, 
the differences in resulting magnitudes for the objects of interest that are 
produced by different competent photometry procedures are a measure 
of the robustness of the magnitude measurements. 

The random error of a single magnitude can be determined from a
comparison of the values obtained by the three different
reduction procedures after zero point and scale difference 
are removed.
The rms. deviation of the triple  
measurements of 250 
stars with $<\!V\!>$ $\sim$ 23.0 and $<\!I\!>$ $\sim$ 22.0 is $0\fm{05}$ 
at this level and increases, of course, towards fainter 
magnitudes.  

\subsection{The Color-Magnitude Diagrams}
\label{CMD}
The observed color-magnitude diagrams (CMDs) obtained with 
ROMAFOT, DoPHOT and HSTphot photometry are presented in 
Fig.~\ref{fig8} -  \ref{fig10}, respectively.




Stars brighter than $V$ = 19.2 and $I$ = 18.3 are saturated and have
therefore not been included in the object list.
The number of well fitted stars with $V$ {\it and\/} 
$I$ is $\sim 9000$ in the case of ROMAFOT, $\sim 38000$ for DoPHOT,
and $\sim 20000$ for HSTphot. These numbers are not directly 
comparable because in each case different lower fitting
limits have been set, which were guided by a subjective 
estimate of what is practical and desirable. 
In addition, the total number of stars in each list
can be altered by
choosing different parameters like goodness-of-fit, 
signal-to-noise ratio, and in the case of HSTphot 
also object-sharpness. 

At brighter magnitudes, for instance $V < 24.5$, the number of stars for
which $V$ and $I$ magnitudes were obtained with the three different
reduction procedures is more comparable. 3666 stars from ROMAFOT stand
against 3469 stars from HSTphot and 6012 stars from DoPHOT.

The reason for the different number of stars is the different
star-searching routines employed. The number of selected stars increases
with the decreasing size of the area to define the "sky". 
DoPHOT and HSTphot employed a particularly small number of pixels for the "sky"
definition in the star-searching routines. In contrast, ROMAFOT uses a 
relatively large "sky" in the star-searching routine. 

The CMDs of ROMAFOT show a number of blue stars at faint levels which
are missing in the other CMDs. This indicates that the ROMAFOT
magnitudes have larger scatter below $V \sim 25$.

The comparison with DoPHOT and HSTphot show that the ROMAFOT
procedure used here is adequate for the discovery 
of the brightest and hence most reliable Cepheids in M\,83.

\subsection{Identification of the variable stars}

The identification of the variable stars is based on the method
by \citet{Lafler:Kinman:65} and is 
described in \citet{Saha:1990}.  The quantities $\Theta$ and the standard
deviation $\sigma(V)$ in function of $V$ over all 34 epochs
were used to identify the variable candidates. 
If P is the
period of a supposed variable star, $m_i$ the measured magnitude at the
ith epoch, and $\overline{m}$ the average over the n values of $m_i$, and 
if the values for $m_i$ are arranged in increasing order of phase, 
then $\Theta$ is defined as: 
\begin{eqnarray}
\label{def_theta}
\Theta(P) \; = \; \frac{\sum_{i=1}^n(m_{i + 1} \; - m_i)^2}{\sum_{i=1}^n
(m_i \; - \overline{m})^2}.
\end{eqnarray}
A minimum in the spectrum of $\Theta$ indicates a possible period. 

We have adopted $\Theta_{\rm min}$, i.e. the lowest smoothed value,
obtained by varying the period between 1 and 570 days. Several hundred
stars were suspected to be variable on the basis of ROMAFOT
photometry. 
The light curves of these possible candidates with reasonable values
of $\Theta$ and $\sigma$ were individually inspected by eye, they were
scrutinized and 12 bona fide Cepheids were retained of which 
10 have also good light curves in $I$. 

{
\begin{deluxetable}{ccccrrcc}
\tablecaption{$V$ and $I$ Magnitudes, Periods, and Positions
of the Selected Cepheids}
\tablewidth{0pt}
\tablehead{
\colhead{Object ID} &
\colhead{$<\!V\!>$} &
\colhead{$<\!I\!>$} &
\colhead{Period [days]} &
\colhead{$X$} & 
\colhead{$Y$} &
\colhead{RA} &
\colhead{DEC}} 
\startdata
C1  &  22.23  &  21.12   &   43.52   & 245.33  & 715.22  & 13:36:55.78 & -29:48:00.67 \\
C2  &  22.61  &  21.59   &   54.92   & 752.38  & 185.02  & 13:36:47.99 & -29:49:46.56 \\
C3  &  23.52  &  22.75   &   28.95   & 544.20  & 449.98  & 13:36:51.18 & -29:48:53.62 \\
C4  &  23.02  &  22.03   &   33.16   & 1577.03 & 863.07  & 13:36:35.33 & -29:47:31.12 \\
C5  &  23.35  &  22.31   &   28.31   & 1283.45 &  39.25  & 13:36:39.84 & -29:50:15.73 \\
C6  &  24.03  &  22.57   &   28.63   & 836.94  & 236.24  & 13:36:46.68 & -29:49:36.37 \\
C7  &  23.22  &  22.27   &   32.55   & 904.89  & 319.27  & 13:36:45.65 & -29:49:19.80 \\
C8  &  23.96  &  23.17   &   14.17   & 1146.40 & 214.52  & 13:36:41.95 & -29:49:40.77 \\
C9  &  24.05  &  22.97   &   12.47   & 652.41  & 161.79  & 13:36:49.54 & -29:49:51.15 \\
C10 &  24.11  &  23.05   &   14.21   & 929.42  & 475.08  & 13:36:45.27 & -29:48:48.64 \\
C11 &  23.42  &  \nodata &   26.22   & 261.55  & 445.71  & 13:36:55.52 & -29:48:54.40 \\   
C12 &  24.34  &  \nodata &   19.25   & 685.64  & 77.02   & 13:36:49.02 & -29:50:08.13 \\   
\enddata
\label{tab2}
\end{deluxetable}
}
The selection criteria whether a star is a Cepheid or not are 
subjective.
The criteria which have been used here are: the quality of the 
light curve in V, the quality of the light curve in I, the phase 
coherence between the $V$ and the $I$ light curves and the shape
of the spectrum of $\Theta$. The curves
are always smoothed to prevent spikes from indicating spurious
periods. A star is more likely to be a Cepheid if the minima 
of $\Theta$ are
broad and a second pronounced minimum exists
at the two-fold period. 

The mean magnitude $<\!V\!>$, i.e. the magnitude of the
phase-weighted intensity average,
was calculated for each Cepheid analog to \citet{Saha:1990} using:
\begin{eqnarray}
\label{phase_weighted_mean}
<\!V\!> \; = \; - 2.5 \; \log_{10} \sum_{i=1}^{n} \; 0.5 \; ( \Phi_{i+1} - \,\Phi_{i-1} ) \, 
10^{-0.4 \,m_i}, 
\end{eqnarray}
where 
n is the number of observations, $m_i$ the magnitude, and $\Phi_i$ 
the phase of the ith observation in order of increasing phase. Intensity
weighted magnitudes can be biased due to missing measurements. 
The phase-weighted intensity mean gives isolated points more
weight than closely spaced ones, which makes it superior to a straight
intensity mean.  

The method to determine the corresponding
mean $<\!I\!>$ magnitude from the few available epochs in $I$ is  
described in \citet*{Labhardt:etal:97}. 
Information on the shape and the amplitude of the complete $V$ 
light curve as well as the typical phase shift between $V$ and $I$ 
are used to derive a value of $<\!I\!>$ from every
single $I$ magnitude:
\begin{eqnarray}
\label{eq_I_mean}
<\!I\!>\;\; =&&\!\ I(\phi_V)  +  [<\!V\!> - V(\phi_V)] + \Delta V \, C_{V \rightarrow I}(\phi), 
\end{eqnarray}
where $\Delta V$ is the $V$ amplitude, $<\!V\!>$ the phase-weighted 
mean $V$ magnitude,
$\phi$ the phase of the light curve and $C_{V\rightarrow I}(\phi)$ the
empirical function for the transformation between $V$ and $I$ 
magnitudes that is tabulated by \citet{Labhardt:etal:97}.
The mean of the individual $<\!I\!>$ magnitudes yields the adopted
value of $<\!I\!>$ and its error.

The light curves of the Cepheid candidates are shown in Fig.~\ref{fig4}.
The candidates C11 and C12 have reasonable light curves in $V$, but
no counterpart in $I$.
Besides those 12 candidates no further Cepheid candidate have been 
accepted. 
Inspection of images (done before period analysis was performed)
indicate that C2, C3, C5, C6,
C8, C9 \& C10 show no evidence of crowding or blending 
whatsoever. We cannot rule out this possibility based on this
inspection for the other objects.
We are providing a deep image of our field which will be available
in the electronic version of the journal. This is better
quality information than images printed on paper. 
The error bars in Fig.~\ref{fig4} in function of apparent magnitude 
have been determined from the reproducibility of artificial stars
of known magnitude. 

\clearpage


\clearpage

The 12 Cepheids are listed in 
Table~\ref{tab2}. Column~1 gives the designation
of the Cepheid, column~2 the period, columns~3\,-\,4 
give the ROMAFOT mean magnitudes and columns~5\,-\,6
their position on the template image V\_17.
The list of Cepheids with $P \, > \, 12^{\rm{d}}$ has no
claim for completeness. The spacing of the observing epochs,
mainly dictated by seeing and weather conditions, may
have made additional Cepheids undetectable
within the surveyed field.  

\section{The Period-Luminosity Relation and the Distance Modulus} 

\subsection{The P-L Relations in $V$ and $I$} 

In a first step we adopt the P-L relation in $V$ and $I$ 
from \citet{Madore:Freedman:1991} as 
\begin{equation}
 M_{V} ~~=~~ -2.76 ~\log P - 1.40~, 
\end{equation} 
\begin{equation}
\label{eq:PL_M_F_I}
 M_{I} ~~=~~ -3.06 ~\log P - 1.81~.  
\end{equation} 
The zero-point of equations (6) and (7) is based on an assumed LMC
modulus of 18.50.

The P-L relations in $V$ and $I$ with the slopes of
equation (6) and (7) are fitted to the 
12 Cepheids with $V$ and the 10 Cepheids with $I$ magnitudes 
in Fig.~\ref{fig11}.
Comparing the P-L relations in apparent magnitudes with 
equations (6) and (7) leads then to (provisional)
apparent moduli $\mu_V$ and $\mu_I$ and their errors 
as shown in Fig.~\ref{fig11} and repeated in Table~\ref{tab3}.


\begin{deluxetable}{lccrc|ccrc|ccrc}
\tablecaption{Individual distance moduli and reddening values 
of Cepheids in M\,83. }
\tablewidth{0pt}
\tablehead{
\colhead{} &
\multicolumn{4}{c}{M/F PL$^{\rm \, a}$} &
\multicolumn{4}{c}{Gal. PL$^{\rm \, b}$} &
\multicolumn{4}{c}{LMC PL$^{\rm \, c}$} \\
\colhead{ID}  &  
\colhead{$\mu_V$} &
\colhead{$\mu_I$} &
\colhead{$E^{\rm \, d}$} &
\colhead{$\mu_0$} &
\colhead{$\mu_V$} &
\colhead{$\mu_I$} &
\colhead{$E^{\rm \, d}$} &
\colhead{$\mu_0$} &
\colhead{$\mu_V$} &
\colhead{$\mu_I$} &
\colhead{$E^{\rm \, d}$} &
\colhead{$\mu_0$}}
\startdata
C1  & 28.15 & 27.94 & 0.21 & 27.64 & 28.20 & 28.03 & 0.17 & 27.78 & 28.05 & 27.83 & 0.22 & 27.52  \\ 
C2  & 28.81 & 28.72 & 0.09 & 28.60 & 28.90 & 28.84 & 0.06 & 28.76 & 28.68 & 28.58 & 0.10 & 28.45  \\   
C3  & 28.95 & 29.03 & -0.08& 29.15 & 28.94 & 29.06 & -0.12& 29.23 & 28.90 & 28.96 & -0.06& 29.05  \\  
C4  & 28.62 & 28.49 & 0.13 & 28.31 & 28.62 & 28.54 & 0.08 & 28.41 & 28.54 & 28.41 & 0.13 & 28.21  \\  
C5  & 28.76 & 28.56 & 0.20 & 28.28 & 28.74 & 28.58 & 0.16 & 28.36 & 28.70 & 28.49 & 0.21 & 28.19  \\ 
C6  & 29.45 & 28.84 & 0.61 & 27.95 & 29.43 & 28.86 & 0.57 & 28.04 & 29.39 & 28.77 & 0.62 & 27.86  \\  
C7  & 28.79 & 28.71 & 0.08 & 28.58 & 28.80 & 28.75 & 0.05 & 28.68 & 28.72 & 28.62 & 0.10 & 28.48  \\  
C8  & 28.54 & 28.50 & 0.04 & 28.45 & 28.40 & 28.42 & -0.02& 28.44 & 28.57 & 28.50 & 0.07 & 28.42  \\  
C9  & 28.47 & 28.13 & 0.34 & 27.64 & 28.32 & 28.03 & 0.29 & 27.61 & 28.52 & 28.15 & 0.37 & 27.62  \\  
C10 & 28.69 & 28.39 & 0.30 & 27.95 & 28.56 & 28.30 & 0.26 & 27.94 & 28.72 & 28.39 & 0.33 & 27.91  \\  
C11 & 28.74 &\nodata&\nodata&\nodata&28.70 &\nodata&\nodata&\nodata&28.69 &\nodata&\nodata&\nodata \\  
C12 & 29.29 &\nodata&\nodata&\nodata&29.20 &\nodata&\nodata&\nodata&29.28 &\nodata&\nodata&\nodata \\  \hline
    & 28.72 & 28.53 & 0.19 & 28.26 & 28.69 & 28.54 & 0.15 & 28.33 & 28.68 & 28.47 & 0.21 & 28.17  \\
&$\pm$0.11&$\pm$0.10&0.06&$\pm$0.15&$\pm$0.11&$\pm$0.11&0.06&$\pm$0.16&$\pm$0.11&$\pm$0.10&0.06&$\pm$0.15 \\
\enddata
\tablenotetext{a}{PL-relation of \citet{Madore:Freedman:1991}}
\tablenotetext{b}{Galactic PL-relation (eq. 8 \& 9)}
\tablenotetext{c}{LMC PL-relation (eq. 10 \& 11)}
\tablenotetext{d}{$E = E(V$$-$$I) = \mu_V - \mu_I$}
\label{tab3}
\end{deluxetable}

\citet{Freedman:etal:01} have suggested, based  
on LMC Cepheids given by \citet{Udalski:etal:99}, 
that the P-L relation in $I$ as given in 
equation~(\ref{eq:PL_M_F_I}) is too steep, and that
consequently all Cepheid distances should be reduced by
$\sim 7$\%. However, the situation
is more complex. 

Based on excellent photometry \citep{Berdnikov:etal:2000} and reddening
values \citep{Fernie:etal:1995} of many hundreds of fundamental-mode Galactic
Cepheids and on corresponding data of even more LMC Cepheids 
\citep{Udalski:etal:99b}, \citet*{Tammann:etal:2003a} 
have shown that the 
period-color (P-C) relations of these two galaxies are distinctly
different and that therefore also their P-L relations must differ
for different wavebands. The new, quite steep Galactic P-L relations are
calibrated by 25 fundamental pulsators in open clusters
and associations 
(Feast 1999; based on a Pleiades zero point of $\mu_0 = 5.61$
[Stello \& Nissen 2001]) {\it and\/} independently
by 28 fundamental pulsators
whose purely physical distances have been derived by 
\citet{Gieren:etal:1998} 
from the Baade-Becker-Wesselink method as revised by
\citet{Barnes:Evans:1976}. 
The adopted calibration agrees also with the HIPPARCOS 
analysis by \citet{Groenewegen:Oudmaijer:2000} to within 
$\sim 0\fm{20} \pm 0\fm{15}$, the latter being brighter. 

The new Galactic P-L relations are given by 
\citet*{Tammann:etal:2003a}:
\begin{equation}
 M_{V} ~~=~~ -3.14 ~\log P - 0.83~, 
\end{equation} 
\begin{equation}
 M_{I} ~~=~~ -3.41 ~\log P - 1.33~.  
\end{equation} 

{\it Preliminary\/} P-L relations of LMC were derived from
650 dereddened fundamental pulsators with good 
photometry by \citet{Udalski:etal:99b}. Assuming again
$(m-M)^{0} = 18.50$, 
\citet{Tammann:etal:2003b} and \citet{Tammann:etal:2001} found for
the LMC Cepheids with $\log P > 1.0$:
\begin{equation}
 M_{V} ~~=~~ -2.48 ~\log P - 1.75~, 
\end{equation} 
\begin{equation}
 M_{I} ~~=~~ -2.82 ~\log P - 2.09~.  
\end{equation} 

There is no question, that the slopes of the P-L relations
in the Galaxy and in the LMC are {\it different\/},
the Galactic slope being steeper for long-period 
Cepheids.

\subsection{The adopted Distance Modulus}

To determine a true distance modulus $(m-M)^0$ from the
apparent Cepheid moduli in $V$ and $I$ the ratio between 
$E(B-V)$ and the absorption in $V$ and $I$ is needed, 
i.e. $A_V$ and $A_I$.
We adopt the following values, that have been
specifically derived for Cepheids
\citep*{Tammann:etal:2003a} 
\begin{eqnarray}
\label{eq:reddening}
R_V &=&  A_V/E(B-V) \, = \, 3.17 \\
R_I &=&  A_I/E(B-V) \; = \, 1.87. 
\end{eqnarray}
The small color dependence of $R$ can be neglected here
because of the restricted range of Cepheid colors. 
The true distance modulus $\mu_0$ becomes then
\begin{equation}
\label{eq:mu_0}
\mu_0 \, = \, 2.44 \, \mu_I - 1.44 \, \mu_V \, . 
\end{equation}

At this point it is not clear which of the three P-L relations
discussed in Section 4.1 should be applied to the M\,83 Cepheids, - the
P-L relations of \citet{Madore:Freedman:1991} as derived from the LMC 
data available at the time, the new P-L relations
of LMC (again at $\mu_{\rm LMC}^0 = 18.50$), or the Galactic P-L
relations whose slopes {\it and} zero point depend on purely 
Galactic data. Therefore all three versions have been applied to the
individual Cepheids of M\,83. The individual apparent moduli 
$\mu_V$ and $\mu_I$ are transformed into a true modulus $\mu_0$
by means of equation~(\ref{eq:mu_0}). The results are shown
in Table~\ref{tab3}.

The three solutions in Table~\ref{tab3} differ by
0$\fm{16}$ at most. 
This near agreement is a fortuitous result, and occurs, 
because the Galactic and 
LMC P-L relations cross over not far from the median period
($28\dm{9}$) of the Cepheids under consideration. 

{\it If\/} metallicity is the main reason for the different P-L
relations in the Galaxy and in the LMC, then the distance modulus 
based on the Galactic P-L relation is more applicable,
because M\,83 with [Fe/H] = 0.3 \citep{Calzetti:etal:1999}
is chemically more comparable
with the Galaxy than with the metal-poor LMC. However,
since we have at present no way of proving that 
metallicity only decides about the slope
of the P-L relations, we adopt a mean modulus of
\begin{equation}
\mu_0 \, = \, 28.25 \pm 0.15 ({\rm statistical \,\, error}) 
\,\, \pm 0.15 \,\, ({\rm systematic \,\, error}),
\end{equation}
which corresponds to $4.5 \pm 0.3 \pm 0.3$ Mpc. 
The statistical error of 0.15 is the standard error of the 
different true distance moduli for each Cepheid in Table~\ref{tab3}.
The estimated systematic error is driven mainly by the difficult photometry
and the non-uniqueness of the P-L relation of Cepheids.

Four of our Cepheids, 
C1, C3, C4 \& C9 can be alleged to be less than fully convincing.
By removing various combinations of these 4 Cepheids from the sample,
we can change the distance modulus to vary between 28.16 to 28.42,
which is consistent with our estimate.

\subsection{Comparison with Previously Published Distances}

The Cepheid distance of M\,83 is important in as much as the galaxy does
not render easily to other methods of distance determinations. An early
distance of 8.9 Mpc, based on the size of the largest H II regions 
\citep{Sandage:Tammann:1974}, was much too large, presumably because its
largest H II regions are relatively small, particularly for an Sc I-II 
galaxy, as it was classified at the time. \citet{deVaucouleurs:1979} 
derived a distance of 3.7 Mpc from several of his distance indicators. 
Pierce's (1994) Tully-Fisher distance of $4.8 \pm 1.0$ Mpc
is unreliable because M\,83 with an inclination of only 34$^0$ is 
not well suited for the method. Adopting a more recent
luminosity class II for M\,83 and an apparent magnitude of
m$^{o,i}_B = 8.08$ (corrected for Galactic and internal absorption;
Sandage \& Tammann 1987) and combining this with Sandage's (2000)
calibration of Sc II galaxies of $M_B = -20.36 \pm 0.68 (H_0 = 50)$,
yields $4.9 \pm 1.8$ Mpc, which is even more insecure because 
of the wide luminosity scatter among Sc II galaxies. A model-dependent 
distance comes from
the expansion parallax of the type II SN 1968L in M\,83 of $4.5 \pm
0.8$ \citep{Schmidt:etal:1994}. Of particular interest is the TRGB 
distance of M\,83 by \citet{Karachentsev:etal:2002a} who found
$\mu_0 = 28.27 (\pm 0.15)$.

\section{Comparison of the Distances of NGC\,5253 and M\,83}

A comparison of the distance of M\,83 with that of NGC\,5253,
which has produced the SNe~Ia 1895B and 1972E, is
interesting because it has been suggested that the two galaxies have 
interacted roughly 1 Gyr ago 
\citep{vandenBergh:1980, Calzetti:etal:1999}. 
In that case they are expected to be
still rather close neighbors. This issue is made even more acute,
given the $0\fm{4}$ difference between 
the distance of NGC 5253 by \citet{Gibson:etal:2000} and the 
original distance by 
the HST Supernovae Consortium (HSTSNC) \citep{Saha:etal:1995,
Tammann:etal:2001}, 
even as they are both based on Cepheids. 

The HSTSNC reduced their HST
observations with two reduction packages: ROMAFOT
as implemented in MIDAS was used at Basel by
L. Labhardt and H. Schwengeler, and the modified DoPHOT
reduction procedure as applied in Baltimore was used by
A. Saha. The resulting magnitudes are in (very) good
agreement, including the $<\!V\!>$ and $<\!I\!>$ magnitudes of 
the five Cepheids in common that
were accepted by HSTSNC as being reliable.

The modulus of NGC 5253, corresponding to $\mu_0 = 28.08$,
was determined 
by the HSTSNC from these 
five excellent Cepheids. The apparent modulus $\mu_V$ 
from the DoPHOT
photometry of {\it seven additional\/} Cepheids is in agreement 
with that from the 
5 `excellent' ones. However, in our judgment, the $I$ magnitudes 
for these 7 are not as reliable: they fall below our adopted
DoPHOT signal-to-noise
threshold for reliable detection. 

One must also appreciate that the 
aberrated HST telescope produced images at the time with diffraction 
structure that could easily result in {\it false detections\/}, 
for which reason 
the detection thresholds had to be kept high. To that was added the 
problem of the very crowded nature of these fields. 
There was therefore good reason to keep 
the selection criterion for acceptable Cepheids very conservative.

While a good many other putative variables were in fact detected both 
in Baltimore and at Basel, they were not considered further for fear 
of polluting with specious objects and erroneous photometry.
If the errors in the apparent moduli in $V$ and $I$ for the 5 
excellent Cepheids are propagated, the formal uncertainty is $\pm 0\fm{28}$.
With this in mind, the path taken by the HSTSNC was to examine the 
{\it differential\/} extinction between Cepheids and the type Ia
Supernovae 1972E in NGC\,5253, 
which is shown to provide tighter constraints on the SNIa calibration.
This bypassed the difficulty of obtaining the distance {\it per se\/} to 
NGC\,5253, when the real goal was to calibrate $M_{V}$ for the 
SNIa.

The HST observations
were re-analyzed by \citet{Gibson:etal:2000}, as part of 
their competing effort by the Mould-Freedman-Kennicutt (MFK) et al. 
group's work to obtain $H_{0}$. They used a different philosophy
of adoption or rejection of candidates.
In their re-reduction of the NGC\,5253 data, \citet{Gibson:etal:2000} 
claimed to find several additional Cepheids not already published by 
\citet{Saha:etal:1995}. These fainter additional objects would not 
have survived the more conservative selection criteria of 
\citet{Saha:etal:1995}. They would have been deemed unusable. 
Of the 7 Cepheids used by Gibson et al., only 2 are in common with 
the 5 excellent ones from \citet{Saha:etal:1995}. In addition 3 objects 
were found by both studies, {\it but were judged 
unusable\/} by \citet{Saha:etal:1995}.
The remaining 2 from \citet{Gibson:etal:2000} were not found in the 
Saha et al. study. The reported photometry from the 5 objects in common 
are in good agreement in both $V$ and $I$ 
\citep[cf.][Table~3]{Gibson:etal:2000},  
even though 3 of them 
were not used by \citet{Saha:etal:1995}. The comparison makes it clear 
{\it that it is not the photometry that is in question\/}, 
as also pointed out by \citet{Gibson:etal:2000}, 
but that the distance derived is sensitive to the {\it sample\/} 
of Cepheids chosen. 

We believe that the conservative selection of the 5 Cepheids 
in \citet{Saha:etal:1995} yields a more reliable sample 
compared to the 7 Cepheids 
used by \citet{Gibson:etal:2000}, particularly since 3 of the latter 7 
were explicitly rejected by \citet{Saha:etal:1995}, and the remaining 
2 were not found by them. In this context it is worth remarking that 
in obtaining distances to the galaxies observed by the MFK et al. 
project, only those Cepheids were used that were deemed worthy by both 
the ALLFRAME based procedure and by the DoPHOT based one. The DoPHOT
procedure used by the MFK et~al. group was identical to the one used
by the HSTSNC. By this reckoning, only 2 Cepheids 
are in the common sample. Using averaged magnitudes from both 
studies for {\it only\/} these 2 common objects, we obtain 
$\mu_{V} = 28.08 \pm 0.18$, 
$\mu_{I} = 28.02 \pm 0.18$, and so $\mu_{0} = 27.93 \pm 0.18$. This result 
is in better agreement with the \citet{Saha:etal:1995} value of 
$28.08 \pm 0.28$  than with the
\citet{Gibson:etal:2000} result of $\mu_{0} = 27.61 \pm 0.11$.
In addition, the difficulties with other 
claims by \citet{Gibson:etal:2000}, 
which we reject, concerning our 
previous photometry of other galaxies in the SNIa calibration 
sample, are discussed at length in \citet{Parodi:etal:2000}. 

In analogy to Table~3 we calculate the individual distances 
of the 5 accepted Cepheids of NGC\,5253 in Table~\ref{tab4}
using again the standard P-L relation of \citet{Madore:Freedman:1991},
and the Galactic and LMC P-L relations by 
\citet*{Tammann:etal:2003a,Tammann:etal:2003b}.
Since NGC\,5253 may be metal-poor, more comparable to
LMC than to the Galaxy, a Cepheid modulus of 
$\mu_{0} = 28.09 \pm 0.25$ is adopted.

Another distance of NGC\,5253 can be obtained from its SN~Ia 1972E.
The apparent magnitude at maximum for this supernovae 
is $m_V^{\rm corr} (\rm max) = 8.49 \pm 0.15$ 
\citep[corrected for Galactic absorption, decline rate, and intrinsic color
according to the prescription in][]{Parodi:etal:2000};
the internal absorption suffered by the SNIa in its host galaxy is
judged to be negligible on the basis of its outlying position
and its color of $(B-V)_{\rm max} = -0.02$. This is bluer,
if anything, than the mean reference color of $(B-V)_{\rm max} = 
-0.01$ of unreddened SNe~Ia \citep[cf.][]{Parodi:etal:2000}.
The mean absolute magnitude
of eight Cepheid-calibrated SNe~Ia (excluding SN~1972E) is 
$M_V^{\rm corr} = -19.47 \pm 0.07$ \citep{Saha:etal:2001b}. 
From this follows a distance
modulus of NGC\,5253 of $\mu_0 = 27.96 \pm 0.19$, where the
statistical error allows for the intrinsic scatter of $\pm 0\fm{11}$
of the corrected absolute magnitude of SNe~Ia (according
to the same authors).
The principal systematic error of this distance determination
comes from the eight calibrating galaxies whose Cepheid distances 
are subject to the current problem of P-L relations 
not being the same in different galaxies.

\begin{center}
\begin{deluxetable}{lccc|ccc|ccc}
\hspace{-3cm}
\tablecaption{Individual distance moduli of Cepheids in NGC\,5253. }
\tablewidth{0pt}
\tablehead{
\colhead{}      &
\colhead{} &
\colhead{M/F PL$^{\rm \, a}$} & 
\colhead{} &
\colhead{} &
\colhead{Gal. PL$^{\rm \, b}$} & 
\colhead{} &
\colhead{} &
\colhead{LMC PL$^{\rm \, c}$} & 
\colhead{} \\
\colhead{ID $^{\rm \, d}$}  &
\colhead{$\mu_V$} &
\colhead{$\mu_I$} &
\colhead{$\mu_0$} &
\colhead{$\mu_V$} &
\colhead{$\mu_I$} &
\colhead{$\mu_0$} &
\colhead{$\mu_V$} &
\colhead{$\mu_I$} &
\colhead{$\mu_0$}}
\startdata
C2-V3  & 28.24 & 28.03 & 27.72 & 28.03 & 27.88 & 27.65 & 28.33 & 28.09 & 27.74  \\ 
C3-V2  & 27.95 & 28.31 & 28.83 & 27.78 & 28.20 & 28.80 & 28.00 & 28.33 & 28.81  \\   
C3-V6  & 28.37 & 28.28 & 28.14 & 28.10 & 28.06 & 28.03 & 28.44 & 28.34 & 28.19  \\  
C4-V2  & 28.08 & 27.79 & 27.37 & 27.94 & 27.70 & 27.36 & 28.11 & 27.79 & 27.34  \\  
C4-V3  & 27.81 & 28.05 & 28.40 & 27.70 & 27.99 & 28.41 & 27.82 & 28.04 & 28.35  \\ \hline
       & 28.09 & 28.09 & 28.09 & 27.91 & 27.96 & 28.04 & 28.14 & 28.11 & 28.09  \\
& $\pm$0.10&$\pm$0.09&$\pm$0.25&$\pm$0.07&$\pm$0.08&$\pm$0.26&$\pm$0.11&$\pm$0.10&$\pm$0.25 \\
\enddata
\tablenotetext{a}{PL-relation of \citet{Madore:Freedman:1991}}
\tablenotetext{b}{Galactic PL-relation (eq. 8 \& 9)}
\tablenotetext{c}{LMC PL-relation (eq. 10 \& 11)}
\tablenotetext{d}{ID as in \citet{Saha:etal:1995}}
\label{tab4}
\end{deluxetable}
\end{center}

The modulus of NGC\,5253 from SN 1972E 
agrees well with that adopted from Table~\ref{tab4},
which is based on the 5 Cepheids by \citet{Saha:etal:1995}.
Our conclusion is that the best weighted distance modulus
is $\mu_0 (\rm{NGC\,5253}) = 28.01 \pm 0.15 \, (4.0 \pm 0.3 
\, {\rm Mpc})$. It should be noted that if the distance
of \citet{Gibson:etal:2000} had been used, SN~1972E
would become the faintest among nine Cepheid-calibrated
SNe~Ia and would be 2.5 $\sigma$ below the mean 
absolute magnitude of the other eight SNe~Ia 
\citep[cf.][]{Saha:etal:2001b}.

The distance difference between M\,83 and NGC\,5253 becomes
then $\Delta \mu = 0.24 \pm 0.21 \, (0.5 \pm 0.4 \, {\rm Mpc})$,
which is not inconsistent, given the errors,
with zero. The projected distance is only 0.15 Mpc.
An interaction 
of the two galaxies roughly 1 Gyr ago is therefore well 
possible. 

The strongest suggestion for a gravitational interaction between
M\,83 and NGC\,5253 comes from the amorphous (Am) 
morphology of the latter. \citet{Hogg:etal:1998} found in an 
objective sample of ten Am galaxies that all have a 
peculiar velocity field and that seven of them have 
nearby companions. One of the three sample galaxies
which they did not assign a companion is NGC\,5253.
The hypothesis of an interaction is supported by
its very unusual gas dynamics 
\citep{Kobulnicky:Skillman:1995}, and by the upper
age limit of $10^8 - 10^9$ years of the stellar
population in the exceptionally large halo of
NGC\,5253 \citep{vandenBergh:1980,
Caldwell:Phillips:1989}. Moreover, also M\,83
shows signs of a post-interaction; the
complex dynamics in its unusually large, lob-sided
HI halo may otherwise be difficult to explain
\citep{Huchtmeier:Bohnenstengel:1981}. 

\section{The M\,83 Group}

\citet{Karachentsev:etal:2002a} have divided the B6 (Cen~A) group
of \citet{Kraan-Korteweg:Tammann:1979} and \citet{Kraan-Korteweg:1986a,
Kraan-Korteweg:1986b}
into two subgroups; one centered on Cen~A, the other on M\,83. 

They list 28 certain or probable members of the Cen~A group.
For seven of them the authors have determined tip of the red giant
branch (TRGB) distances, their mean value being 
$\mu^0 = 27.80 \pm 0.04 \, (3.63 \pm 0.07$ Mpc). 
For Cen~A proper the authors give a TRGB distance of 
$\mu^0 = 27.81 (\pm 0.15)$ and \citet{Rejkuba:2002} 
$\mu^0 = 27.99 (\pm 0.10)$ from the TRGB and Miras.
For 14 members of the Cen~A group \citet{Karachentsev:etal:2002a}
also list $v_{\rm LG}$ velocities (their Table~2), corrected to the 
centroid of the Local Group, and with a mean value of 
$v_{\rm LG} = 293 \pm 24 \kms (\sigma_{\rm v} = 90 \kms)$. 
The correction from heliocentric velocities $v_{\sun}$ to 
$v_{\rm LG}$ is large in the direction of Centaurus and 
sensitive to the adopted solar apex solution. The authors 
have adopted the solution of \citet{Karachentsev:Makarov:1996}.
We prefer the solution of \citet{Yahil:Tammann:Sandage:1977},
which has the advantage of being independent of any adopted
distances and which excludes companion galaxies whose
orbital motion may deteriorate the solution. In this case
one obtains $\Delta v_{\rm LG} = -275$ (instead of $-245$)
$\kms$ and hence $v_{\rm LG} = 263 \pm 31 \kms$ (allowing
for an additional error of $ 20 \kms$ in $\Delta v_{\rm LG}$)
for the radial velocity of the Cen~A group.

The M\,83 group with 11 members (including now NGC\,5253) is clearly
more distant. \citet{Karachentsev:etal:2002a} give TRGB
distances for five dwarf members. Their mean distance is
$\mu^0 = 28.57 \, (\pm0.15)$. This compares reasonably with 
the adopted distances of M\,83 ($\mu_0 = 28.25 \pm 0.15$
from Cepheids) and NGC\,5253 ($\mu_0 = 28.01 \pm 0.15$
from Cepheids, and SN 1972E). Combining the distances of the
dwarfs, M\,83, and NGC\,5253 gives a mean distance of the
M\,83 group of $\mu_0 = 28.28 \pm 0.10 \, (4.5 \pm 0.2 {\rm Mpc})$,
further than the Cen A group.

Redshifts of 10 members of the M\,83 group are listed in the NASA
Extragalactic Database. Their mean recession velocity is 
$v_{\sun} = 494 \pm 37 \kms$ or - with $\Delta v_{\rm LG} = -245 
(\pm 20) \kms$ \citep*{Yahil:Tammann:Sandage:1977} 
- $v_{\rm LG} = 249 \pm 42 \kms$. The additional 
correction to this value for a Virgocentric infall model
is very small because M\,83 lies close to the surface where
the velocity components (in the radial direction of the observer)
of the respective Virgocentric velocity vectors of the Local
Group and of M\,83 nearly cancel
\citep[cf.][]{Kraan-Korteweg:1986a,Kraan-Korteweg:1986b}.
If the local infall vector is assumed to be 220 $\kms$, 
$\Delta v_{220}$ becomes $+3 \kms$ taking the M\,83 group 
at 4.5 Mpc and the Virgo Cluster at 21.5 Mpc.
An observer at the centroid of the Local Group would therefore
observe a recession velocity of the M\,83 group of 
$252 \pm 42 \kms$ if there was no disturbance from the 
Virgo complex. 


The distances of the Cen~A and M~83 groups as seen from the 
centroid of the Local Group - assumed to lie between the Galaxy
and M~31 at 2/3 of the M~31 distance \citep{Sandage:1986} -
is 0.5 Mpc larger than seen from the Sun, hence $r$ (LG $-$ Cen~A)
$ = 4.13 \pm 0.1$ Mpc and $r$ (LG $-$ M~83) $ = 5.0 \pm 0.2$ Mpc. 
In case of pure Hubble flow and a cosmic value of $H_0$ = 60
\citep{Parodi:etal:2000, Saha:etal:2001a, Tammann:etal:2001}
the predicted group velocities would be $v_{\rm LG}$(Cen~A)
$ = 248 \pm 6$ and $v_{\rm LG}$(M~83) $ = 300 \pm 12 \kms$.
These values differ from the observed ones by only $-15 \pm 32$
and $+48 \pm 44 \kms$. Thus the deviations from pure Hubble
flow remain within the measurement errors.

\section{Conclusions}

Twelve Cepheids with periods 12$^{\rm d} <  P < 
55^{\rm d}$ were found in NGC\,5236 (M\,83) with the 
8.2\,m 
ANTU (UT1) telescope of the VLT. This shows that the telescope
in its present configuration 
can be used for work on Cepheids out to $\sim$ 
5 Mpc. The advantage over the much used WFPC2 of HST
is the wider field and the better
sampling of stellar images. Disadvantages are the 
enhanced problem of crowding and the larger number 
of epochs required for period determinations, because an
optimized epoch distribution is hampered by external observing
conditions. In spite of this, the total exposure time 
needed for M\,83 (60000 sec) is somewhat shorter than
for a typical Cepheid distance with the WFPC2 (85000 sec).

The photometry was carried out with ROMAFOT. One epoch in 
$V$ and $I$ were independently reduced with DoPHOT and
HSTphot. The photometric zero points agree to $\pm 0\fm{03}$
at $V = 20.0$ and $I = 19.0$. The three magnitude systems
show, however, a small difference in the photometric scales
leading to mean magnitude differences of $\la 0\fm{1}$
at $I$ = 22.5; the scale error at $V$ = 23.5 is only about 
half this value. These differences are explained 
by the different philosophies how to treat the galaxy 
background. 
This problem is inherent to all photometries of
faint stars which are seen against a bright background.

The distance of M\,83 was derived from the ten Cepheids  
which have good light curves in $V$ and $I$ using three
different P-L relations, i.e. the steep Galactic
P-L relation, calibrated through Cepheids in open clusters
{\it and\/} with Baade-Becker-Wesselink distances, and 
two versions of the LMC
P-L relation based both on an assumed modulus of 
$(m-M)^0_{\rm LMC} = 18.50$. The resulting mean distance
of M\,83 is found to be $(m-M)^0 = 28.25 \pm 0.15 \,
(4.5 \pm 0.3 \, {\rm Mpc})$. 

The distance of NGC\,5253, based on its Cepheids and 
SNIa 1972E is rediscussed. The resulting modulus
of $(m-M)^0 = 28.01 \pm 0.15 \, (4.0 \pm 0.3 \, {\rm Mpc})$
confirms the earlier Cepheid distance by 
\citet{Saha:etal:1995} and disagrees with that of
\citet{Gibson:etal:2000}.

The hardly significant difference in radial distance
between M\,83 and NGC\,5253 and the small {\it projected\/}
distance (0.15 Mpc) make it possible
that the two galaxies have interacted in the past,
which may be the origin of the unusual amorphous type of
NGC\,5253. 

The importance of the distances of M\,83 and NGC\,5253, 
strengthening the mean distance of the M\,83 group of
$(m-M)^{0} = 28.28 \pm 0.10$ based also on TRGB distances
of the dwarf members of \citet{Karachentsev:etal:2002a}, 
is the continuing proof of the quietness of the local Hubble flow 
just outside the Local Group. Although the group is not the 
closest of the very local galaxies to the zero velocity surface 
that separates the beginning of the expansion field from the bound 
galaxies of the Local Group \citep{Sandage:1986}, it nevertheless is 
close enough to be important for the eventual mapping of
the position of this surface at the Local Group boundary. 

At the time of \citet{Sandage:1986}
the Im dwarf galaxies Leo A and Pegasus Dw (DDO 
216) were considered to be well beyond the 
Local Group. Their low velocities reduced to the Local Group 
centroid are $-32 \, {\rm km s}^{-1}$ and $+ 62 \, {\rm km s}^{-1}$ 
respectively. These are 
very low for their large distances of 1.6 Mpc and 2.5 Mpc, assumed 
on the basis of the extant literature of 1986. It was 
this circumstance that gave considerable weight to the apparent 
detection of a deceleration due to the Local Group.
\citet{Aparicio:1994} for 
Pegasus and 
\citet{Tolstoy:etal:1998} and \citet{Dolphin:etal:2002} for Leo A 
have shown that they are actually
members of the Local Group.
\citet{Aparicio:1994} and \citet{Dolphin:etal:2002} derived both a 
distance modulus of about (m - M)$^0 = 24.5$, whereas 
\citet{Tolstoy:etal:1998} derived (m - M)$^0 = 24.2$.
{\it Hence, the evidence of deceleration based on Leo A and Pegasus 
has disappeared\/}. 

The question of deceleration will be discussed in a forthcoming
paper using galaxies near the edge of the Local Group such as
Sextans~A and B, the Antlia dwarf (LGS3), GR~8, NGC~300, as
well as the M~81/NGC~2403 and the IC~342 groups. {\it The emerging
picture is that the expansion is already well in progress
at a distance of $\sim$ 1.5 Mpc\/} 
\citep[cf.][]{Ekholm:etal:2001, Karachentsev:etal:2002b}. 
The M\,83 group at 4.5 Mpc and
with a corrected velocity of $v_{\rm LG} = 249 {\rm kms}^{-1}$ fits well
into this picture.

The {\it absence\/} now of detectable deceleration 
of the velocity field outside the Local 
Group in the 
presence of a Local Group mass of $2 \times 10^{12} M_{\sun}$ 
\citep{Sandage:1986, vandenBergh:1999, 
Evans:etal:2000} or even $4.9 \times 10^{12} M_{\sun}$, 
as suggested by its dynamical history 
\citep{Lynden-Bell:1999}, demands an 
explanation, as does the obviously {\it small dispersion\/} of 
the random velocities of the very local galaxies. This {\it coldness\/} 
of the velocity field in the presence of the clearly lumpy 
distribution of the visible matter has been a puzzle since 1972 
\citep*{Sandage:etal:1972}.

      A modern suggestion that the total force field is nearly 
homogeneous (smooth) due to the dominance everywhere of an all 
pervasive cosmological constant, diluting any lumpy gravity field 
of the clustered matter, has been put forward now by many, perhaps 
the first being \citet{Chernin:Teerikorpi:Baryshev:2000}. These 
considerations make a continual precision mapping of the local 
velocity field on a scale of 10 Mpc even more crucial. The work 
on M\,83 here is a contribution to this central problem.

\smallskip\noindent
{\bf Acknowledgments}

We thank the VLT team who performed so successfully the
service mode observations. 
F.\,T. and G.\,A.\,T. thank the Swiss National Science Foundation
and the Swiss PRODEX programme for financial support. F.\,T. thanks
the National Optical Astronomy Observatories (NOAO) for their
hospitality. 




\pagebreak

\figcaption{The DSS image of the 20 x 20 arcmin field centered 
at the position of M 83. North is up and East is to the left.
The insert shows the 6.8 x 6.8 arcmin FORS1 
field of view northwest of the center of M\,83.
\label{fig1}}

\figcaption{VLT FORS1 $V$ image of the target field northwest of the 
center of M 83 which illustrates the high spatial 
resolution. Marked objects are the Cepheids in
Table\ref{tab2}.
\label{fig2}}

\figcaption{Comparison of $V$ (left) and $I$ (right) magnitudes derived from 
measurements obtained with ROMAFOT and DoPHOT and a matching
radius of 1 pixel.
\label{fig5}}

\figcaption{Same as  Fig.~\ref{fig5} but for magnitudes derived from 
measurements obtained with ROMAFOT and HSTphot.
\label{fig6}}

\figcaption{Same as Fig.~\ref{fig5} but for magnitudes derived from 
measurements obtained with DoPHOT and HSTphot.
\label{fig7}}

\figcaption{$V-I$ versus $V$ color-magnitude diagram obtained with ROMAFOT.
\label{fig8}}

\figcaption{$V-I$ versus $V$ color-magnitude diagram obtained with DoPHOT.
\label{fig9}}

\figcaption{$V-I$ versus $V$ color-magnitude diagram obtained with a
modified version of HSTphot.
\label{fig10}}

\figcaption{The light curves of 10 Cepheids in $V$ (filled
circles) and $I$ (open circles). Two additional Cepheids, 
C11 and C12, have no measurements in $I$. 
\label{fig4}}

\figcaption{Period-luminosity relation of M\,83
in $V$ (top) and $I$ (bottom)
for all 12 Cepheids in $V$ and all 10 Cepheids 
in $I$. The solid lines represent the best fit with the canonical 
slope of $-2.76$ in $V$ and $-3.06$ in $I$.
The Cepheids C11 and C12 with no $I$ magnitude
are plotted with open symbols.
The dashed lines account for an adopted intrinsic width of the
instability strip of $\pm 0\fm{4}$ in $V$ and $\pm 0\fm{32}$ 
in $I$.
\label{fig11}}

\end{document}